# Jupiter's third largest and longest-lived oval: Color changes and dynamics

N. Barrado-Izagirre[1], J. Legarreta[2], A. Sánchez-Lavega[1], S. Pérez-Hoyos[1], R. Hueso[1], P. Iñurrigarro, J. F. Rojas[1], I. Mendikoa[1], I. Ordoñez-Etxeberria[1] and the IOPW Team[3]

[1]*Departamento de Física Aplicada I, Escuela de Ingeniería de Bilbao, Universidad del País Vasco UPV/EHU, Plaza Torres Quevedo 1, 48013 Bilbao, Spain*

[2]*Departamento de Ingeniería de Sistemas y Automática, Escuela de Ingeniería de Bilbao, Universidad del País Vasco UPV/EHU, Plaza Torres Quevedo 1, 48013 Bilbao, Spain*

[3] *http://www.pvol2.ehu.eus/*

**Highlights:**

- **We study a long-lived color changing anticyclone in Jupiter from 2012 to 2019.**
- **Tangential velocity and vorticity are below half the value of GRS and BA.**
- **Despite being a weak vortex it has survived for years after mergers and disturbances.**

*Corresponding author:* Naiara Barrado-Izagirre, naiara.barrado@ehu.eus

**Abstract**

The transition region between the North Equatorial Band (NEBn) and North Tropical Zone (NTrZ) in Jupiter is home to convective storms, systems of cyclones and anticyclones and atmospheric waves. Zonal winds are weak but have a strong latitudinal shear allowing the formation of cyclones (typically dark) and anticyclones (typically white) that remain close in latitude. A large anticyclone formed in the year 2006 at planetographic latitude 19ºN and persists since then after a complex dynamic history, being possibly the third longest-lived oval in the planet after Jupiter's Great Red Spot and oval BA. This anticyclone has experienced close interactions with other ovals, merging with another oval in February 2013; it has also experienced color changes, from white to red (September 2013) and back to white with an external red ring (2015-2016). The oval survived the effects of the closely located North Temperate Belt Disturbance, which occurred in October 2016 and fully covered the oval, rendering it unobservable for a short time. When it became visible again at its expected longitude from its previous longitudinal track, it reappeared as a white large oval keeping this color and the same morphology since 2017 at least until the onset of the new convective disturbance in Jupiter's North Temperate Belt in August 2020. Here we describe the historic evolution of the properties of this oval. We use JunoCam and Hubble Space Telescope (HST) images to measure its size obtaining a mean value of $(10,500\pm1,000) \times (5,800\pm600)$ km$^2$ and its internal rotation finding a value of $-(2\pm1)\cdot10^{-5}$s$^{-1}$ for its mean relative vorticity. We also used HST and PlanetCam-UPV/EHU multi-wavelength observations to characterize its color changes and Junocam images to unveil its detailed structure. The color and the altitude-opacity indices show that the oval is higher and has redder clouds than its environment but has lower cloud tops than other large ovals like the GRS, and it is less red than the GRS and oval BA. We show that in spite of the dramatic environmental changes suffered by the oval during all these years, its main characteristics are stable in time and therefore must be related with the atmospheric dynamics below the observable cloud decks.

# 1. Introduction

The ubiquitous presence of vortices at cloud level is one of the most important characteristics of the meteorology of the giant planets along with the zonal winds organized in a multiple jet system. Jupiter is the most prolific planet in showing a great variety of closed circulation vortices (Rogers, 1995). Vortices with sizes above 2,000 km are observed at all latitudes of the Jovian disk except very close to the Equator (Mac Low and Ingersoll, 1986; Morales-Juberías et al., 2002a, 2002b; Li et al., 2004). Recent observations obtained by the Juno mission show also stable cyclones close to both poles (Orton et al., 2017; Adriani et al., 2018, 2020; Tabata-Vakili et al. 2020). The vortices can be visually distinguished by their reflectivity contrast with respect to adjacent clouds and by their shape, showing an oval form that encircles a region of closed or nearly-closed vorticity. The surrounding cloud patterns make them appear as "bright" or "dark" ovals with sizes ranging from one hundred kilometers to 40,000 km, the maximum size measured at the end of the XIX century for the largest oval in Jupiter, the well-known Great Red Spot (GRS) (Simon et al., 2018a).

Vortices are classified according to their relative vorticity as cyclones or anticyclones. In the Northern Hemisphere, cyclones show anti-clockwise rotation while anticyclones rotate clockwise. Anticyclones appear in a great number and a variety of sizes in the anticyclonic domains of the Jovian zonal wind profile, being more stable than cyclones (Vasavada and Showman, 2005), except at polar latitudes where cyclones are the stable vortices (Adriani et al., 2018, 2020). The most apparent and well known vortices, due to their longevity and size, are the Great Red Spot (GRS, planetographic mean latitude 22ºS) and oval BA (planetographic mean latitude 33ºS) (Figure 1), both anticyclones.

The essential properties to understand a vortex are its vorticity distribution and its relation with the environment flow shear (Dowling and Ingersoll, 1989; Marcus, 1993). Other aspects that can be important in Jovian vortices are their color, possible changes in time, and interactions with other vortices, which may include mergers (constructive interactions) or destructive interactions with eddies. In the giant planets, where the mechanisms powering the zonal winds are mostly unknown, the interaction between vortices, eddies and jets are one of the mechanisms proposed to have an important role in forcing and maintaining the zonal jets (Ingersoll et al. 2004).

The first vorticity measurements in Jupiter's atmosphere were obtained for the GRS using ground-based photographic observations (Reese & Smith, 1968; Hess, 1969). These were later much improved with precise measurements from the Voyager 1 and 2 flybys (Smith et al., 1979a, 1979b) including accurate measurements of the detailed flow field (Sada et al. 1996). The Voyagers also obtained precise measurements of the local vorticity for White Oval BC (at 33ºS) (Mitchell et al., 1981) and for a cyclone "barge" at 16ºN (Hatzes et al., 1981). Images obtained by the Galileo obiter allowed to measure the wind field of the GRS (Choi et al., 2007) and a few smaller vortices (Vasavada et al., 1998; Simon et al., 1998). Currently, HST images can also be used to retrieve the internal flow field of the largest vortices such as the GRS or oval BA (e.g. Hueso et al., 2009; Wong et al. 2011) and JunoCam has provided data with enough spatial resolution and temporal separation to measure the wind field of the GRS (Sánchez-Lavega et al. 2018) and polar vortices (Tabataba-Vakili et al. 2020).

When two vortices of the same vorticity type closely interact they can merge (when they are of similar sizes), or if they have very different sizes the smaller one might get absorbed totally or partially. Vortex mergers occur in different areas of the planet with the best well-known example being the chain of large vortex mergers in 1998 and 2000 that resulted in oval BA (Sánchez-Lavega 1999, 2001). The new white oval BA turned red in August 2005 with a very similar shade to that of the GRS (Naeye, 2006 and Simon-Miller et al. 2006). Extensive dynamic studies (García-Melendo et al., 2009; Hueso et al. 2009; Wong et al. 2011) did not find dynamical differences linked to color. According to radiative transfer analysis models, the color change resulted from the diffusion of a colored compound that interacted with the solar photons at the upper levels of the oval (Pérez-Hoyos et al., 2009). Partial absorptions have been observed in Jupiter's Great Red Spot interactions with large ovals (Sánchez-Lavega et al. 1998; 2021).

Color changes are relatively common in Jupiter's atmosphere. Color changes in the red coloration of tropical anticyclones have been described in Sánchez-Lavega et al. (2013). Their analysis of tropical vortices concluded that the vertical structure and dynamics of the anticyclones are not the causes of their coloration, and they propose that the red chromophore forms when the background material is stirred and exposed to ultraviolet radiation or mixed with other chemical compounds inside the vortex. In addition, planetary-scale disturbances in the Jovian atmosphere can modify the zonal albedo pattern of the planet as for example in the cycle of the South Equatorial Belt (SEB) with convective Disturbances, large-scale color changes and Fades (Sánchez-Lavega and Gómez, 1996; Fletcher et al. 2011, Pérez-Hoyos et al. 2012). The North Temperate Belt (NTB) also experiences this kind of event leading to the entire band becoming totally disturbed (Rogers, 1995, Sánchez-Lavega and Gomez, 1996, Sanchez-Lavega et al., 2008). The NTB Disturbances (NTBD) start with the outbreak of one or more convective plumes seen as bright spots moving with a velocity slightly faster than the zonal wind that interacts with the surrounding cloud patterns altering them and forming turbulence in their wake. The last NTBD developed in October 2016 with an outburst of four plumes (Sánchez-Lavega et al. 2017) that disturbed the entire latitudinal band and led to the formation of a very different reddish band (Pérez-Hoyos et al., 2020).

In this work, we have followed the evolution of a long-lived anticyclone located in the boundary between the North Tropical Zone (NTrZ) and the North Equatorial Belt (NEB) at 19º N planetographic latitude (Figures 1 and 2). This oval has been observed continuously since 2008, but there are evidences of earlier observations (Rogers, 2013). The oval is interesting because of its large size and longevity (third after the GRS and oval BA), its color changes between white and red, and its interactions with close vortices including a major vortex merger in February 2013, and by the interaction with the NTBD in 2012, 2016 and 2020. In the following, we will refer to this anticyclone as NTrZ-Oval (NTrO).

The area where this vortex is located is also interesting, as this region is home to the formation of several anticyclonic white ovals, cyclonic dark barges and a few cases of small red spots (Beebe and Hockey 1986; Sánchez-Lavega and Quesada, 1988; Rogers, 1995). It is also the location of wave systems that interact with the ovals and might be caused by them (Simon et al., 2018b). Our goal is to understand how the environmental changes affect a long-lived oval, trying to discern which of its properties can be taken as fundamental and which other properties are just incidental. Among the general properties of a vortex we will study morphology, color, zonal drift and vorticity.

We present here a full report of the observed characteristics of the NTrO (excluding the interactions with the 2020 NTBD taking place in Jupiter at the time of this writing). Throughout this paper we will use planetographic latitude (Sánchez-Lavega, 2011) and System III longitudes (Davies et al. 1986). This paper is structured in the following way. We detail the observations used in Section 2. We detail the historical evolution of the morphology of the oval, its wind field and color evolution from 2012 to 2019 in section 3. Finally, we present our conclusions in section 4.

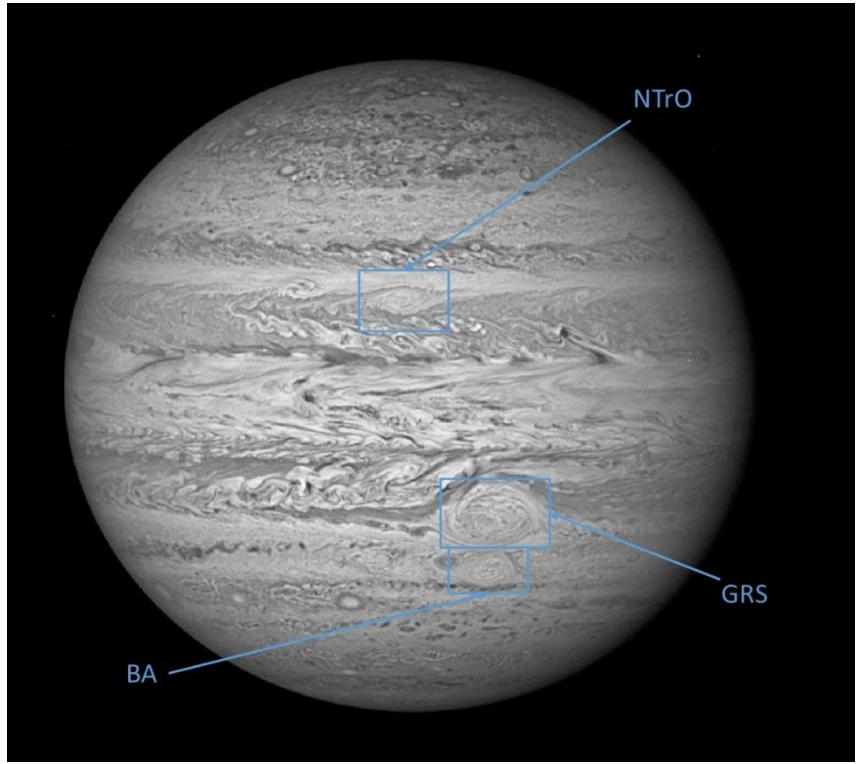

*Figure 1: The three longest-lived and largest Jupiter ovals: GRS, oval BA and the NTrO.* Observation from HST on 20[th] of September, 2012 in F763M filter.

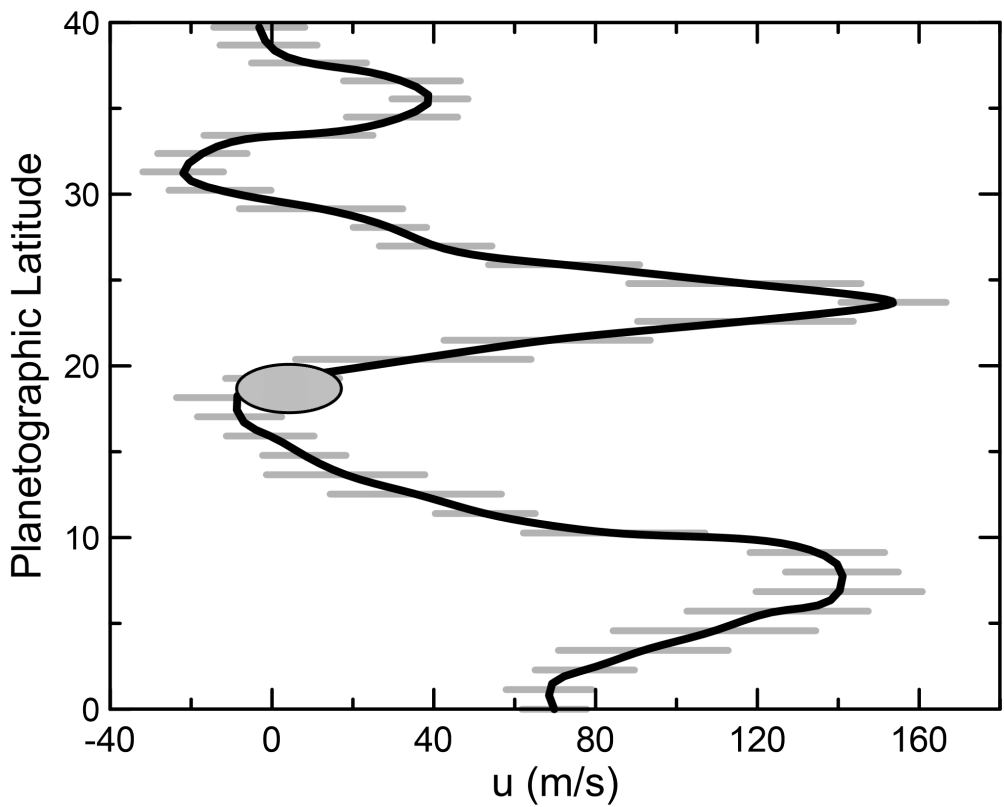

*Figure 2: Zonal wind profile of Jupiter and location of the NTrO.* The zonal wind profile is from Barrado-Izagirre et al., (2013) and the ellipse shows the location and approximate size of the NTrO within it.

## 2. Observations

This work is based on the analysis of several data sets. These include: (i) images acquired with ground-based telescopes using lucky-imaging techniques that allow high-spatial resolution observations using small telescopes (Mousis et al., 2014) and available in public databases such as PVOL (Hueso et al., 2010a, 2018); (ii) photometrically calibrated observations with the PlanetCam instrument (Mendikoa et al., 2016, 2017); (iii) HST observations when available and retrieved from the HST archive. In addition, we used an observation of the oval obtained at very high spatial resolution by the JunoCam instrument (Hansen et al., 2014) on the Juno mission. Figure 3 shows representative examples of these observations (except for the JunoCam image that will be presented later). These sets of images are very different from each other but allow us to obtain a nearly complete temporal coverage for the period of interest. The observations used are summarized in Table 1. We here detail the main characteristics of these data sets.

The images from small telescopes were obtained by amateur astronomers who contribute with their observations to the Planetary Virtual Observatory and Laboratory (PVOL) database at https://pvol2.ehu.eus. This database started as a database of observations of the giant planets from the International Outer Planets Watch (IOPW) collaboration (Hueso et al., 2010a) and currently encompasses images of all solar system planets (Hueso et al., 2018). PVOL stores currently more than 33,000 images of Jupiter acquired since the year 2000 with contributors from more than 30 countries. The quality and effective spatial resolution of these images depend on several parameters: planet apparent size (which depends on the time of observations with respect to opposition date), atmospheric conditions (seeing), telescope diameter, planet elevation with respect to the horizon, camera used and details of the image processing. The greatest strength of these observations is their temporal resolution with tens of images per day available in the days around Jupiter's opposition. The images are generally high-pass processed versions of the original data, with a large fraction of them being RGB colour composites. However, they cannot generally be used for radiative transfer analysis or colour characterization. We also participate as data suppliers with observations acquired with a 28-cm telescope from the Aula EspaZio facility at the School of Engineers of the University of the Basque Country UPV/EHU (Sánchez-Lavega et al. 2014) and from a robotic 35-cm telescope from our university situated at Calar Alto observatory.

We also regularly run observation campaigns (typically 2 campaigns of 4 nights per year) with the PlanetCam-UPV/EHU instrument (Mendikoa et al. 2016) at the 2.2m telescope in Calar Alto observatory. PlanetCam is a fast dual camera acquiring images in the visible (380 nm- 1.0 µm) and SWIR (1.0-1.7 µm) that uses the "lucky imaging" method to obtain high-resolution images (Law et al., 2006). Observations are calibrated in absolute reflectivity by observing standard reference stars (Mendikoa et al., 2017). PlanetCam allows observations in narrow-filter images including filters in the methane absorption bands and near-ultraviolet, which can then be used to perform the color index calculations later described in subsection 3.4.

We analyzed six sets of images acquired by HST using the Wide Field Camera 3 (WFC3) and retrieved from the public archive (see Table 1). The spatial resolution of these images is on average a factor 10 higher (0.04"/pixel) than most typical images in the PVOL database. This translates in effective spatial resolutions at Jupiter tropical latitudes of 150 km/pixel approximately. Unfortunately, HST images correspond to epochs in which the oval appeared white, as there are no HST observations of Jupiter when the oval was red. We used images acquired on 19-20 September 2012 to calculate the internal wind field in the oval and images obtained on January 2015, February 2016, February and April 2017, and April 2018 to measure the internal wind field and also to evaluate color indices of different features.

All the images presented so far need to be navigated, i.e. the longitude and latitude planetary coordinates of each pixel (x, y) of the image must be computed. To navigate HST and ground-based observations, we used the software LAIA, based in the VICAR code (Video Image Communication And Retrieval, Duxbury and Jensen, 1994.; see García-Melendo & Sánchez-Lavega 2001 for details).

Finally, JunoCam observed the NTrZ oval at three perijoves: PJ9 (2017 October 24), PJ16 (2018 October 29) and PJ21 (2019 July 21). In PJ16, observations did not fully cover the oval, but in PJ9, and especially in PJ21, they covered the oval in high-resolution with the best images reaching effective spatial resolutions of 25 km/pixel at the oval. We transformed an original JunoCam image to a cylindrically projected map using Juno's trajectory information from SPICE kernels and the Integrated Software for Imagers and Spectrometers (ISIS3) software of the U.S. Geological Survey.

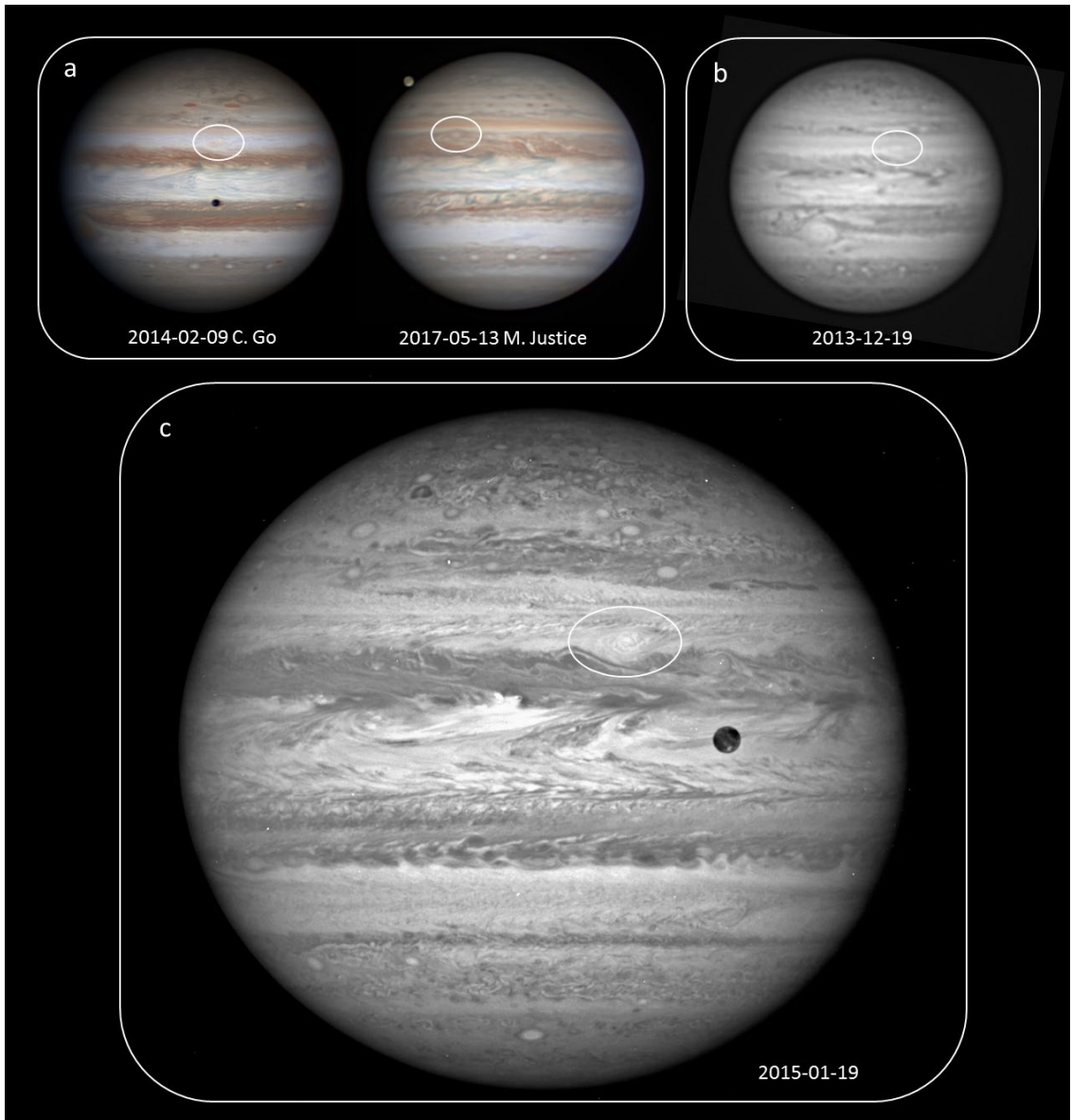

**Figure 3: Examples of images used in this work**. *(a) Observations from amateur astronomers available in the PVOL database. Individual observers are indicated in the figure. (b) PlanetCam observation in a Johnson I filter showing a photometric (unprocessed) image. (c) HST (F658N).*

| Instrument | Date | Number of images | Filters |
|---|---|---|---|
| **HST** | 2012/09/20 | 6 | F275W, F763M |
| | 2015/01/19 | 8 | F658N, F502N, FQ889N, F275W |
| | 2016/02/9-10 | 10 | F631N, F658N, F502N, FQ889N, F275W |
| | 2016/12/11 | 4 | F631N |
| | 2017/01/11 | 4 | F631N |
| | 2017/04/03 | 10 | F631N, F658N, F502N, FQ889N, F275W |
| | 2018/04/17 | 4 | F631N |
| **IOPW** | From 2012 to 2019 | < 900 | RGB, MT |
| **PLANETCAM** | 2013/12 | 12 | B, R, M3, UV |
| **JUNO** | 2019/07/21 | 1 | RGB |

*Table 1: Summary of the HST, PVOL and Juno observations used in this work. Most of the PVOL images are RGB composites (wavelength range from~400 to 700 nm) or IR (685 or 700 high-pass).*

## 3. Analysis

### 3.1. History and morphology

The NTrO has been present in the planet since 2008 and possibly before that (Rogers, 2013). Until 2012 it was one of the many different ovals present in the NEB-NTrZ region and was known as white oval Z following the convention of naming different spots with letters. However, in April 2012 a convective disturbance in the NTB developed. This disturbance started close to Jupiter's solar conjunction and there are only a few images that recorded this event (Rogers and Adamoli, 2019). As a result of this disturbance, the oval changed its drift velocity starting to move relative to the chain of ovals at close latitudes. Afterwards, in February 2013, the NTrO merged with another white oval, named A. The latter was somewhat slower as it was located at 19ºN while Z was at 19.5ºN, where the winds are a little bit faster. The merger led to a very similar white oval (see Figures 4 and 5) that continued the same velocity tendency as it had before as oval Z and was widely known in the amateur community as White Oval Z (WOZ).

Some months later, in September 2013, a dramatic change took place: the oval turned red (Figure 4) and looked brighter in the methane band filter images as we will discuss in section 3.4. It stayed red for more than a year, and was sometimes named Red Oval Z (ROZ), but suddenly, in February 2015, it dimmed at methane band filter images and recovered its initial white coloration except for a pinkish ring.

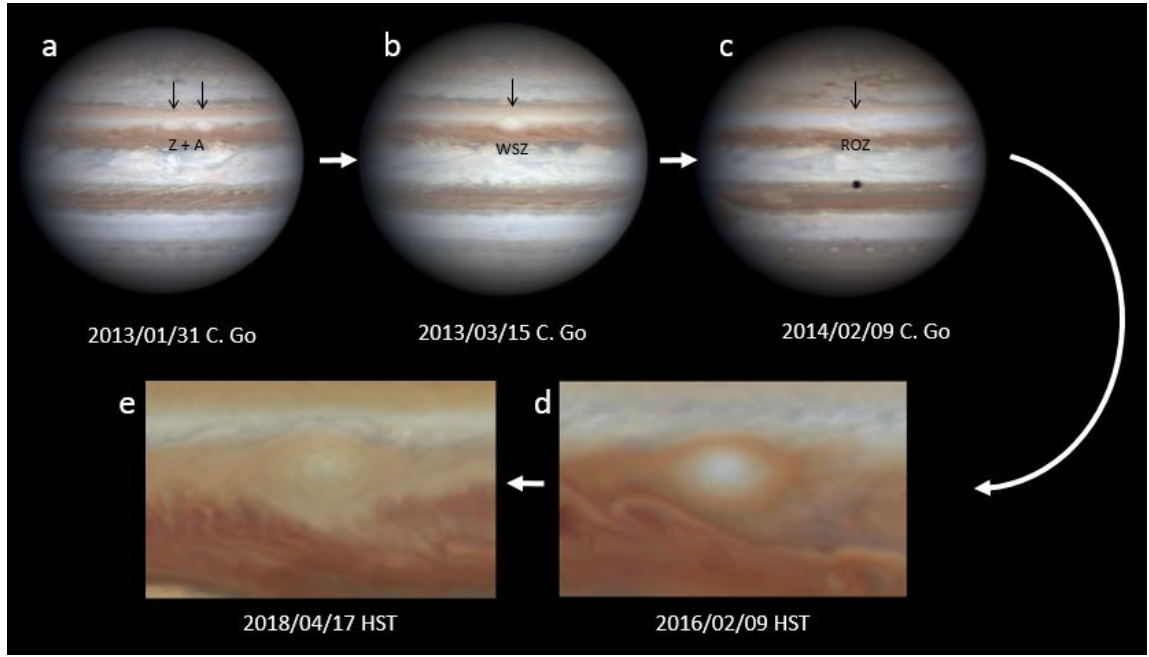

**Figure 4. Evolution of the oval and its aspect in different periods.** *(a) The two ovals Z and A that merged to form the white oval; (b) Immediate result of the merger (White Spot Z); (c) A sample of the oval during its red phase (Red Oval Z); (d) Transitional phase with a red/pink ring; (e) Current state of the oval.*

It remained with that visual aspect until October 2016, when an outburst of four plumes in the NTB gave rise to the 2016 NTBD (Sánchez-Lavega et al., 2017), again during the solar conjunction. When the disturbance ceased, the morphology of the whole NTB had changed. In December 2016, HST observations revealed the persistent presence of the NTrO, but later images in January and February 2017 did not clearly show the oval, as the latitudes where the oval was became very blurry and the oval apparently disappeared (Figure 5). A careful track of the oval in PVOL images made it possible to distinguish the oval from the background and follow its position until December 2017 even though the NTrZ and NTB were totally disturbed. In January 2019, and due to the proximity to solar conjunction, the oval was not distinguishable in the few images available in the amateur record. However, on February 18th it appeared again at its expected longitude of 250º (system III). By that date many new ovals had appeared in the NTrZ possibly as a consequence of the 2016 NTBD. The new ovals were both cyclones and anticyclones, with the NTrO being the largest and most conspicuous of all these features.

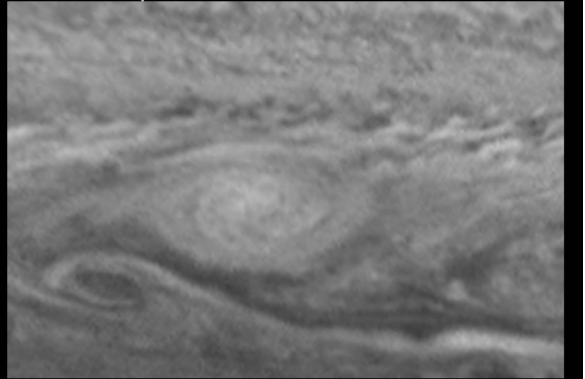
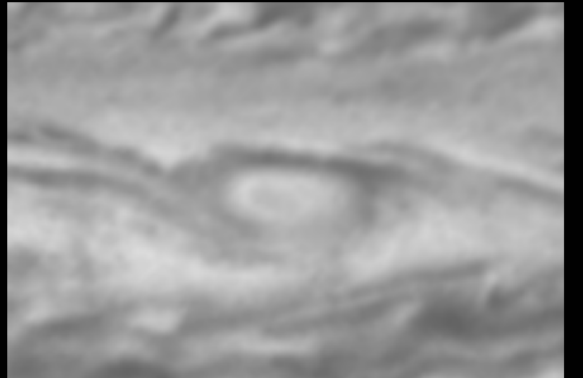
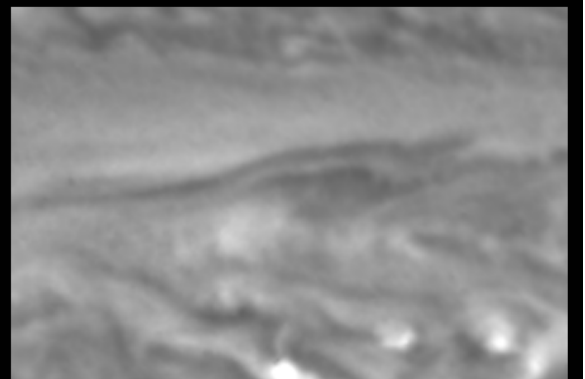
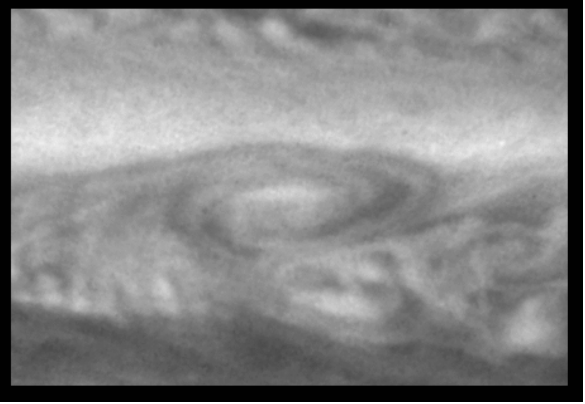

*Figure 5: NTrO in detail before and after the 2016 NTBD event.* *All images from HST in F658N filter.*

Juno observed the oval on different orbits, but generally under poor observing conditions. However, Juno passed above the NTrO on its perijove 21 on 21 July, 2019 and obtained images of the anticyclone at very high spatial resolution. Figure 6 shows a cylindrical map projection based on the Junocam image JNCR_2019202_21C00021 with an effective spatial resolution above the oval of ~25 km. The oval appears as a large-scale feature with a white core and a reddish ring that extends into the environment with fuzzy boundaries. The mean size of the reddish ring is an ellipse with a mean size of (10,500±1,000) x (5,200±400) km² and the inner white region has a mean size of (5,800±200) x (4,500±200) km². Here measurements and errors come from the statistical analysis of different size estimations of the oval after fitting ellipses to its diffuse boundaries. Bright and dark patches in the inner white core have typical sizes of 210±45 km. The morphology of the north and south branches of the oval are different, possibly related with the very different meridional shear of the environment zonal winds, which is much stronger in the northern hemisphere. Small-scale bright features (~150 km) at 21.2º planetographic latitude (read on the right axis of Figure 6), in the transition between the outer red ring and the environment white clouds. This is a region of strong meridional shear of the zonal wind with the relative vorticity of the mean zonal wind being $-\partial u/\partial y = (4.0 \pm 0.8) \times 10^{-5}$ s$^{-1}$. Similar-sized features in the north side of the white oval are located at 20º. The transition in the south branch of the oval from the reddish ring to the white environment at 16.5º latitude occurs in a region of low wind shear with $-\partial u/\partial y = (0.6 \pm 0.8) \times 10^{-5}$ s$^{-1}$. The winds and vortex circulation cannot be directly measured on the Junocam images due to the small time-separation between consecutive images (2 minutes).

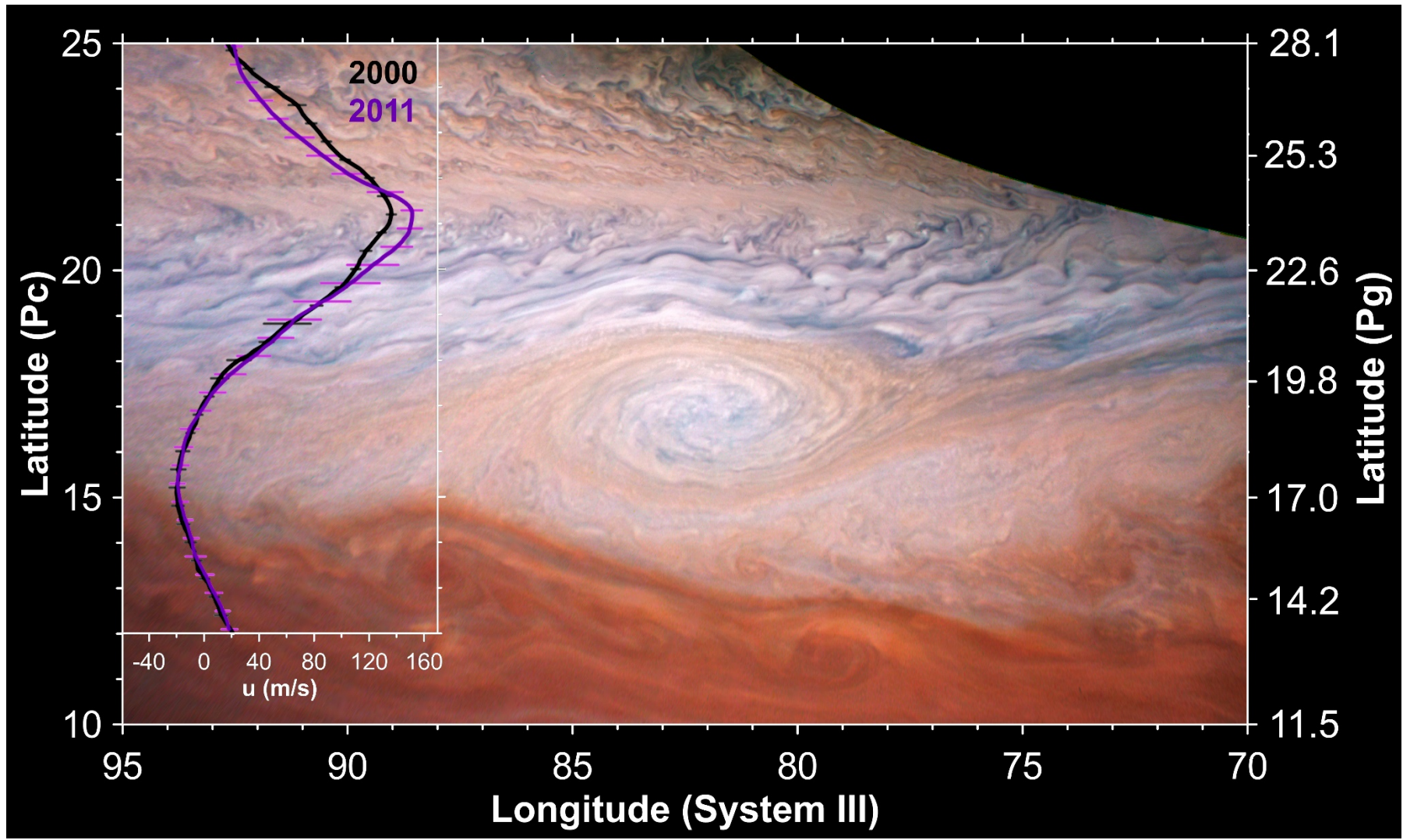

*Figure 6. The White Tropical Oval and its environment from JunoCam observations on June 21, 2019.* The image was originally projected in planetocentric coordinates. The inset shows zonal winds as measured during the Cassini flyby (Porco et al. 2003) in black or in 2011 (Barrado-Izagirre et al., 2012) in purple with only minor changes in this latitude range.

## 3.2. Drift velocity and tracking of the oval

The NTrO is located in the boundary between the North Tropical Zone (NTrZ) and the North Equatorial Band (NEB) of Jupiter at 19º N planetographic latitude. This is an anticyclonic area where the zonal velocity is near 0 m/s (Figure 2), so the oval should be quasi-stationary. However, the wind shear in that area is so pronounced that a small change of less than a degree in latitude leads to a large change in zonal velocity. Moreover, south of its latitude, the local wind produces strong cyclonic vorticity and a small change in its latitude could result in the oval destruction by the interaction with the wind. The longevity of the NTrO in this area, with such a high wind shear (see Figure 2), requires a strong inner circulation. Large and long-lived stable vortices like the GRS and oval BA share also a position in a strongly sheared jet.

We have measured the drift velocity of the NTrO tracking its position since August 2011 until November 2019. Measurements over more than 300 images are shown in Figure 7. Changes in the slope of the plot represent changes in the drift rate of the oval, i.e. its zonal wind speed. Such changes are probably related to small changes in latitude. Given errors are obtained from the standard deviation of the individual measurements.

At the beginning of the tracking period, there is a noticeable velocity change probably related with the NTBD that took place in April 2012. Although only a few images show the White Oval Z in this period, the long time difference between them makes it possible to measure the velocity change that goes from 0.2 ± 0.2 m/s before the NTBD to 13.3 ± 0.5 m/s afterwards.

At the end of 2012, White Oval Z approached a similar oval at a very close latitude, (White Oval A in Figure 2). The tracking of this feature is also shown in Figure 7. White Oval Z was at 19.7 ± 0.6º moving at a velocity of 11.5 ± 0.5 m/s, while oval A was located at 19.1 ± 0.4 º and moved at a velocity of 5.3 ± 0.7 m/s. These velocities correspond to those of the zonal wind profile at their respective mean latitude. Their longitudinal distance in December 2012 was 35º and after two months, they converged and merged into a unique vortex, NTrO that followed the drift tendency of Z (Figure 6). Before and after this merger both vortices were white and showed no color changes.

After the merger, the drift velocity from February 2013 to September 2013 was about 6.5 m/s, while from September 2013 to November 2014 it suffered a deceleration and the drift velocity was around 2.9 m/s and not as uniform as in the first period, when the linear fit was almost perfect. We also note that these changes follow very well the shape of the wind profile of the area because the latitude of the oval also changed very slightly. The color change happened at that very same time. From August 2014 to December 2016 the oval was in a nearly stationary situation, with its latitude very close to 19º N again, with its drift rate matching the zonal wind profile at its latitude.

In October 2016, an outburst of convective plumes in the NTB disturbed the NTB and NTrZ (Sánchez-Lavega et al., 2017). The visual properties of the oval did not change in this period but in the tracking presented in Figure 6 we observe that some days after the eruption of the plumes the track velocity of the oval changed substantially, increasing from being nearly stationary to a velocity of 5.54 ± 0.07 m/s to the East. This behavior is equivalent to the one that had been observed 7 years before at the similar 2012 eruption. In both cases, the NTrO moved in latitude from 19º to 19.5º leading to a region of the zonal wind profile with higher velocities.

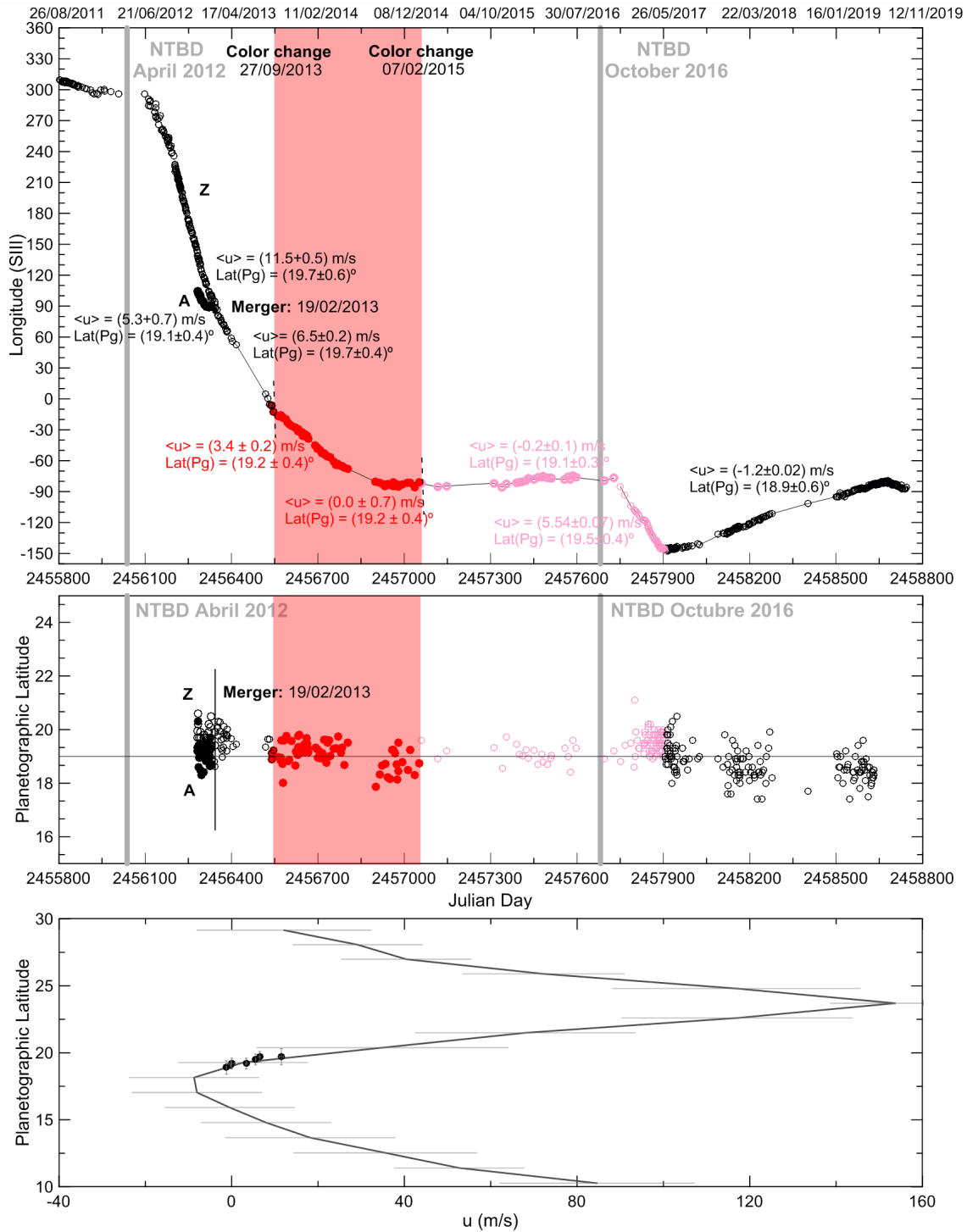

Figure 7: *Tracking of the NTrO oval from August 2011 to November 2019.*
*Top panel: Longitudes. Middle panel: Latitudes. In black, the measurements of the oval when it was white, red dots when the oval looks red and pink circles for situation with a white center encircled by a pinkish ring. The two NTB disturbances occurred during this period are marked with grey areas. Bottom panel: The position of the oval over the wind profile.*

### 3.3. Rotation velocity and vorticity

We selected HST images of the NTrO to measure the internal flow field in the oval. We selected images from September 2012 before the merger happened, when the oval was white, and from January 2015 and February 2016 when the oval was white after its red phase and still had a pink ring surrounding it, prior to the 2016 NTBD. We selected images from January 2017, April 2017 and April 2018 to characterize its properties after the NTBD.

We used the PICV (Particle Image Correlation Velocimetry) 2-dimensional image correlation software (Hueso et al. 2009) to analyze image pairs and study motions in the NTrO. We selected images of the same area obtained with a typical time separation of 10 hours and we navigated and projected them into cylindrical coordinates. The software allows selecting regions of the images where image correlation is used to identify cloud tracers and their motions. Because of the limited resolution of the images, each measurement was "validated" by a human operator that examined the cloud feature identified by the software in both images and its correlation map. We also manually measured features in different data sets to evaluate the effectiveness of the correlator for this particular oval, as the size and contrast of the cloud features may affect the results.

In both cases, zonal $u$ and meridional $v$ velocities are calculated from the displacements in longitude $\lambda$ and planetocentric latitude $\theta$ respectively. $u$ and $v$ are given by the expressions:

$$u = -\frac{\pi}{180} R(\theta) \cos\theta \frac{\Delta\lambda}{\Delta t} \qquad (1)$$

$$v = \frac{\pi}{180} R(\theta) \frac{\Delta\theta}{\Delta t} \qquad (2)$$

where $R(\theta)$ is the local radius at a latitude $\theta$ of the selected detail and is given by

$$R(\theta) = \frac{R_e R_p}{\sqrt{R_e^2 \sin^2\theta + R_p^2 \cos^2\theta}} \qquad (3)$$

where $R_e$ is the equatorial radius (71.492 km) and $R_p$ the polar one (66.854 km) both at 1 bar level.

Motions of cloud features inside an oval follow elliptic trajectories around its center. We used the methodology presented in Legarreta and Sánchez-Lavega (2005) to retrieve the tangential component of the velocity ($V_T$) and the normal velocity ($V_N$):

$$V_T = -u \sin \chi + v \cos \chi \qquad (4)$$

$$V_N = u \cos \chi + v \sin \chi \qquad (5)$$

where $\chi = (a^2/b^2) \tan \phi$, being $\phi$ the polar angle for the detail position with respect to the center of the ellipse and, $a$ and $b$ the mayor and minor semiaxis of the oval ellipse.

Wind results from the correlation software are shown in Figure 8 for the White Oval Z during its initial period in September 2012 before the merger with White Oval A, and from a much later period in 2017. The maximum tangential velocity was found to be concentrated along a peripheral "ring-like" area, so we have calculated the mean tangential velocity for the vortex ($<V_T>$) as the average of the tangential velocities measured at the external ring of the vortex. We found nearly null values of the normal component as in previous studies of jovian vortices (Mitchell et al. 1981). However, HST resolution is not enough to evaluate the radial values and to test three-dimensional models of vortices such as the overturning vortex model proposed by Mahdinia et al. (2017). The values of the mean tangential velocity for these two years and for equivalent data in 2015, 2016 and 2018 are summarized in table 2. The errors for $<V_T>$ are obtained from the standard deviation of the individual measurements along the peripheral area. These results were also checked by manually cloud tracking some details in the oval in good agreement with the correlation results. Unfortunately, we do not have high-resolution images in the time period when the oval was red.

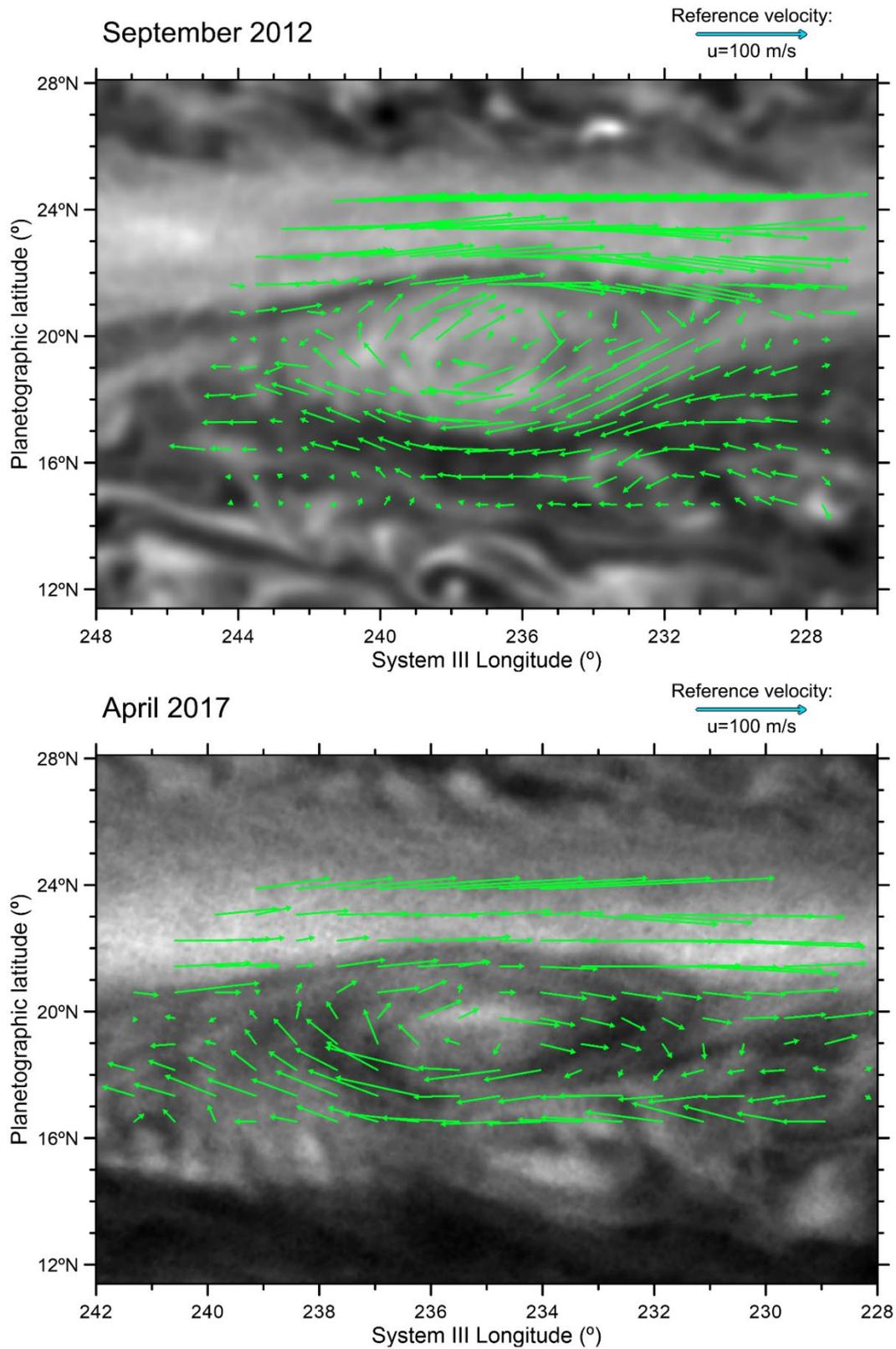

*Figure 8: Examples of measured wind field in NTrO in 2012 and 2017.* These results were obtained with the image correlation software resulting in a dense network of measurements. These measurements were later interpolated over a regular grid.

Once the wind field in the vortex is known we can measure its mean relative vorticity (i.e. the local vertical component of the curl of the two dimensional wind field) using the circulation theorem assuming elliptical trajectories. We follow Sánchez-Lavega et al. (1998). The mean vertical component of the vorticity in a closed contour $L$ for a velocity field in an area $A$ is given by:

$$\zeta = \frac{1}{A} \oint_L \vec{u} \, \vec{dl} \tag{6}$$

From observational measurements, the velocity along the integration contour is constant within uncertainties and we can approximate:

$$\zeta = \frac{1}{A} \oint_L V_T \, dl \approx \frac{1}{A} V_T \oint_L dl = \frac{V_T}{A} L_e \tag{7}$$

where $L_e$ is the perimeter of the ellipse described by the vortex with a semimajor axis $a$ and eccentricity $e$:

$$L_e = 4a \int_0^{\pi/2} \sqrt{1 - e^2 \sin^2 \phi} \, d\phi \approx 2\pi a \left\{ 1 - \left(\frac{1}{2}\right)^2 e^2 - \left(\frac{1 \cdot 3}{2 \cdot 4}\right)^2 \frac{e^4}{3} - O(e^6) \right\} \tag{8}$$

To obtain the error for the relative vorticity we use the quadratic sum of errors from the different parameter measured:

$$\Delta \zeta = \sqrt{\left(\left|\frac{\partial \zeta}{\partial V_T}\right| \Delta V_T\right)^2 + \left(\left|\frac{\partial \zeta}{\partial a}\right| \Delta a\right)^2 + \left(\left|\frac{\partial \zeta}{\partial b}\right| \Delta b\right)^2} \tag{9}$$

Substituting our measurements for the NTrO we obtain values of relative vorticity summarized in Table 2.

*Table 2: NTrO's measured values for different years.*

| Year | a (º) | b (º) | $V_T$ (m/s) | $\zeta$ (s$^{-1}$) |
|---|---|---|---|---|
| 2012 | 4.9 | 2.4 | -60±30 | (-3±1)·10$^{-5}$ |
| 2015 | 4.4 | 2.9 | -35±20 | (-1.6±0.9)·10$^{-5}$ |
| 2016 | 3.2 | 2.1 | -33±10 | (-2.1±0.7)·10$^{-5}$ |
| 2017 | 4.9 | 2.1 | -34±20 | (-2±1)·10$^{-5}$ |
| 2018 | 4.0 | 3.0 | -40±20 | (-1.9±0.9)·10$^{-5}$ |
| 2019 (*) | 4.0 | 2.1 | - | - |

**Notes:** *a is half the east-west extension of the oval, while b is half the north-south extension, $V_T$ is the tangential velocity of the external contour of the oval and $\zeta$ is the relative vorticity measured with eq. 7. Errors associated to this measurement are obtained with eq. 9. (*) Measurements obtained over the JunoCam image.*

The average size of the oval on HST images is 10,500 km zonally (the same value as in JunoCam) and 6,200 km meridionally (slightly larger than in the JunoCam image). As it can be seen in Table 2, this size was not constant and had a minimum value in 2016. This gives an average eccentricity of 0.8±0.1, more elongated in comparison with other long-lived ovals as GRS or BA (Simon et al, 2018a and Perez-Hoyos et al. 2009). The relative vorticity of NTrO ranges from $-1.6 \cdot 10^{-5}$ to $-3 \cdot 10^{-5} s^{-1}$, with an average value of about $-2 \cdot 10^{-5} s^{-1}$. This is half the value of the GRS and BA, around $5 \cdot 10^{-5} s^{-1}$ for the former in the last four years (Simon et al., 2018a) and $4 \cdot 10^{-5} s^{-1}$ for the latter in 2000 (Pérez-Hoyos et al., 2009). Comparing with the ambient vorticity, GRS and BA are in an area with a 5 to 10 times lower value ($1.5 \cdot 10^{-5} s^{-1}$ and $0.4 \cdot 10^{-5} s^{-1}$, respectively) but with a higher Coriolis parameter ($12 \cdot 10^{-5} s^{-1}$ and $19 \cdot 10^{-5} s^{-1}$, respectively). However, the relative vorticity of the zonal jets at the latitudinal location of the NTrO, $\sim -(2.3 \pm 0.8) \cdot 10^{-5} s^{-1}$, is very similar to the relative vorticity of the oval $-(2 \pm 1) \cdot 10^{-5} s^{-1}$, again with a higher Coriolis parameter of $11 \cdot 10^{-5} s^{-1}$. This suggests that the vortex is sustained by the zonal jets confining the vortex.

### 3.4. NTrO in different wavelengths

During the epoch in which the oval was white, around 2012, it was not clearly discernible in images in the methane absorption band available in the PVOL database. However, in September 2013 it was more reflecting than its surroundings in the methane absorption band images indicating higher clouds than its environment and its previous structure. This change was accompanied by a change in color from white to red in colored images and the oval remained bright in methane while it was red (Figure 9), sharing this characteristic with other large anticyclones in Jupiter such as the GRS and the oval BA. Unfortunately, these color and brightness changes observed in images acquired by amateur astronomers cannot be quantified because absolutely calibrated versions of the images are not possible.

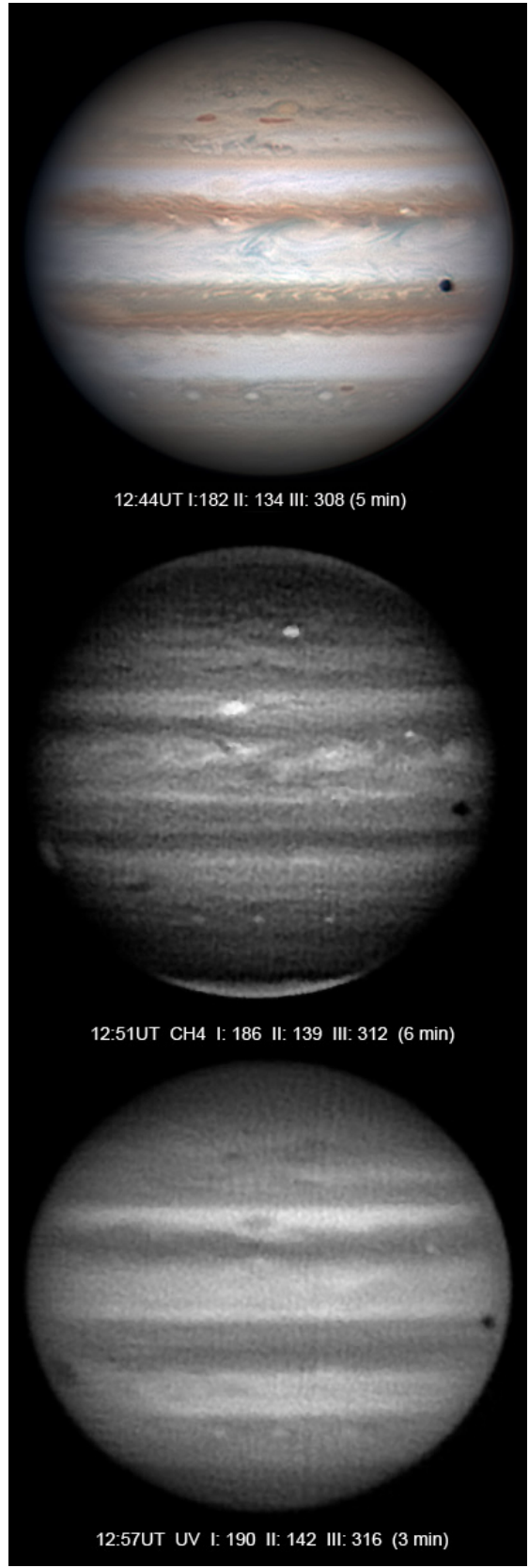

***Figure 9: NTrO in different wavelengths in 2014.*** *Images in RGB, the methane absorption band at 890 nm, and in ultraviolet acquired on February 2nd, 2014 by Christopher Go. Note the faint red color of the oval accompanied by its bright aspect in methane and dark in UV.*

The HST did not observe Jupiter when the NTrO was red. However, there are sets of images obtained later on. Figure 10 shows some examples of the color and brightness evolution from 2015 to 2018. In these images, when compared with other large ovals in the planet, such as the GRS (West et al., 2004) or oval BA (e.g. Pérez-Hoyos et al., 2009), the NTrO looks not as dark in the UV, nor as bright in the 890 nm methane band filter as the other ovals do. This fact suggests that the vortex cloud tops were not as elevated with respect to its surroundings as those of the GRS or oval BA, at least during this period of time. In such a case, the lack of Rayleigh scattering at short wavelengths would produce a darker image, while the presence of high-altitude scatterers would prevent methane absorption and result in a brighter aspect on the methane band filter.

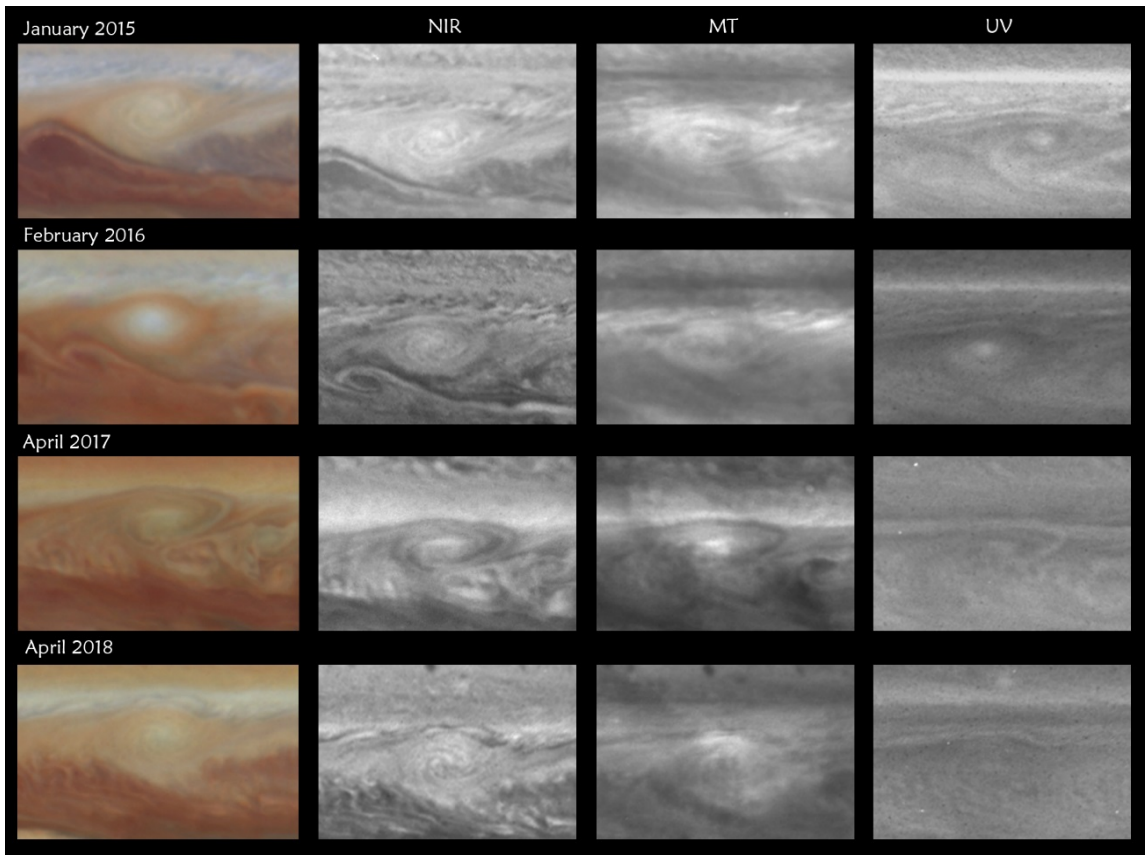

*Figure 10: Appearance of NTrO in different filters and time periods.*

Different techniques have been used to quantitatively characterize colors in Jupiter and guide the search for the chromophore species on the Jovian atmosphere and the relation with the spatial distribution of colors. Principal Component Analysis (PCA) has been a popular technique in this regard (Simon-Miller et al. 2001a; 2001b; Dyudina et al. 2001; Strycker et al. 2011; Ordoñez-Etxebarria et al. 2016). Here we try a more simple approach used by Sánchez-Lavega et al. (2013) that consists of a relative photometry measurement of the features of interest and the calculation of two photometric indices to characterize and compare the color and cloud properties of selected Jovian features. This approach has been also applied to the overall behavior of the planet (Ordoñez-Etxebarria et al., 2016) and for the NTB following its 2016 disturbance (Pérez-Hoyos et al., 2020).

We employ the indexes introduced by Sánchez-Lavega et al. (2013); the color index (CI) and the altitude-opacity index (AOI) based on four filters from the HST observations:

$$CI_k = \frac{R_k(\lambda=410 \text{ nm})}{R_k(\lambda=673 \text{ nm})} \quad (9)$$

$$AOI_k = \frac{R_k(\lambda=890 \text{ nm})}{R_k(\lambda=255 \text{ nm})} \quad (10)$$

Here $R_k$ is the relative reflectivity value of a feature at a given bandpass, relative to a uniform area in the Equatorial Zone. We used the software PLIA (Hueso et al. 2010b) to retrieve the reflectivity from uncalibrated *DN* (Digital Number) counts of the oval taking as a reference for normalization the EZ area close to the longitude where the oval is located. This way relative reflectivity highlights the spectral differences between features. So the relative reflectivity is given by:

$$R_k = \frac{<DN(\lambda)>_k / \mu_{0k}}{<DN(\lambda)>_r / \mu_{0r}} \quad (11)$$

where *k* represents the value for the oval and *r* is used for the reference area. $\mu_{ok}$ and $\mu_{or}$ are the cosines of the illumination angles, which are always as similar as possible for a given image. We run our analysis on HST and PlanetCam images of Jupiter whose brightness distribution and limb darkening have been partially compensated following a Lambert correction. This is enough for the purpose of our analysis as the vortex and the reference are close enough.

Color Indices defined using filters in wavelengths similar to those here have been used in the past by other authors with similar results (de Pater et al., 2010, Wong et al. 2011 and Ordoñez-Etxeberria et al. 2016). For the 2015-2018 HST observations we make use of the F502N (502 nm) and the F658N (658 nm) for the CI and FQ889N and F275W (275nm) for the AOI. On the other hand, for the PlanetCam (2013) images we used B (445 nm) and R (658 nm) for CI and M3 (890 nm) and U (380 nm, shortest wavelength visible from ground-based telescopes) for AOI. For HST images we used calibrated photometric images provided by the HST pipeline. For PlanetCam images, we used unprocessed stacks of the original frames after dark subtraction and flat-corrections, but without any contrast enhancement or image processing.

The color index CI is sensitive to the chromophore because of the spectral slope of the reflectivity in Jupiter between blue and red wavelengths (Karkoschka, 1998). Sánchez-Lavega et al. (2013) found that in Jupiter, the CI index runs from about 0.5 for red features (GRS, RO) to about 1.1 for the "white" features. The altitude-opacity index, AOI, is sensitive to the upper haze altitude optical depth and its ultraviolet opacity. They found values around 0.8 for features with low altitude and UV optically thin upper haze (cyclones) and values around 1.5 for high altitude and thick hazes (anticyclones).

Figure 11 presents our results for NTrO. These quantitative measurements confirm that the oval was red but not so much as the GRS or oval BA in its period of red color. Additionally, in the 2015 and 2016 period, the oval became white with a pink ring around it but in 2017 and 2018 the ring disappeared and does not show up in visual inspection of the images or in our color indices.

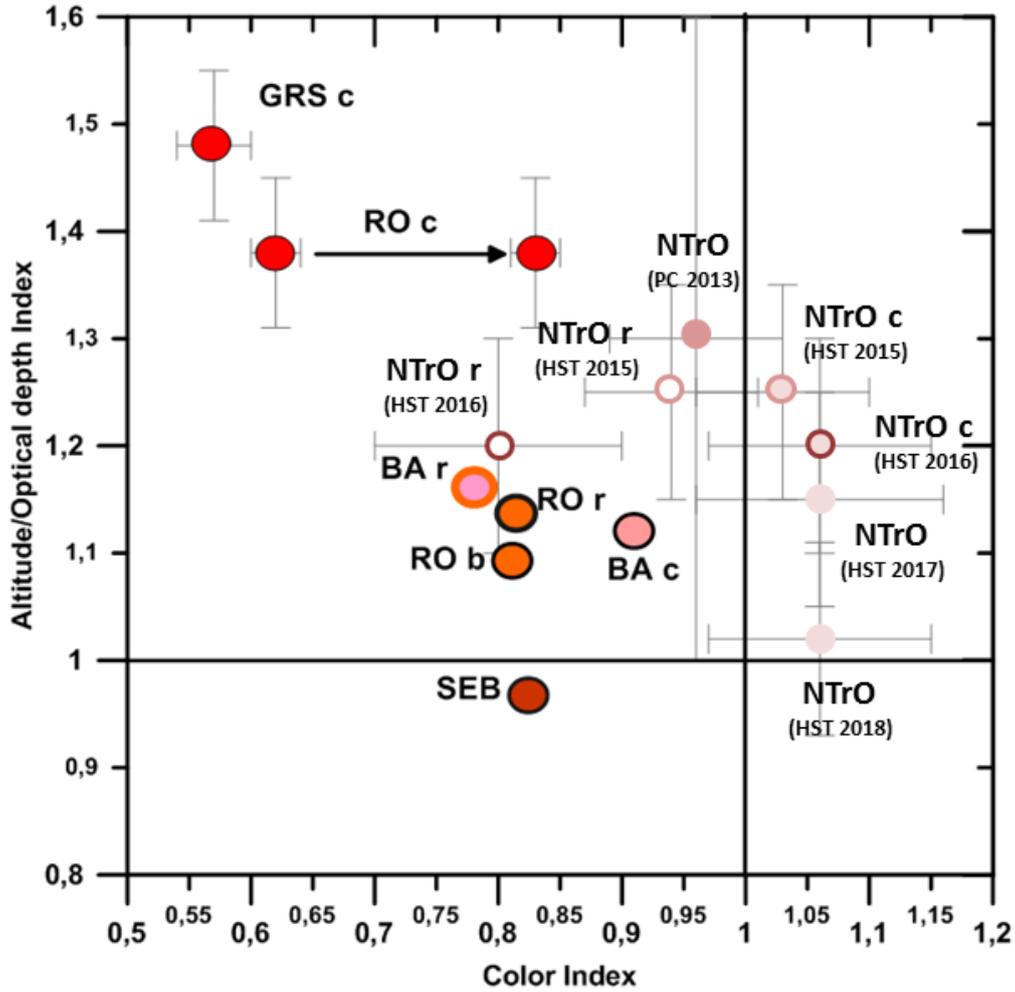

*Figure 11: Scatter plot of color (CI) and altitude opacity (AOI) indexes. We show CI and AOI for different jovian regions studied by Sánchez-Lavega et al. (2013) adding our data for the NTrO. The features from Sánchez-Lavega et al. (2013) are GRS center (GRSc), a Red Oval center before and after its interaction with GRS-BA (ROc), RO background (ROb), oval BA center (BAc) and oval BA ring (BAr). All our measurements for the NTrO are labelled with either HST or PC (PlanetCam) and the year of the observations. Sublabels c indicate "center" and r "ring"*

## 4. Discussion and conclusions

We have studied the evolution of a long-lived tropical anticyclone in Jupiter's atmosphere using a combination of ground-based observations, HST images and observations from JunoCam on Jupiter. This anticyclone, the NTrO, experienced significant changes in its color and drift rate, possibly triggered by the eruptions of the North Temperate

Belt in 2012 and 2016, and is currently the third largest and longest-lived closed atmospheric system in Jupiter's atmosphere. The main conclusions of this work are:

- The NTrO has existed at least since 2008 but it is probably older than that, so it is one of the longest living ovals observed in Jupiter. It is half the size of the second largest oval in Jupiter, oval BA. As oval BA, it has experienced vortex mergers and color changes accompanied by changes in its vertical cloud structure and upper cloud content .

- The oval here studied is an anticyclone with a mean vorticity value of $-(2\pm1)\cdot10^{-5}s^{-1}$, which is very similar, within the error, to the vorticity of the zonal jet at the main latitude of the oval with a vorticity value of $\sim-(2.3\pm0.8)\cdot10^{-5}s^{-1}$, contrarily to GRS and BA which have higher vorticity values than their corresponding ambient vorticity. This suggests that NTrO must be sustained by the zonal jets confining it. Small changes in latitude seem to have produced large changes in its drift rate without significantly altering its size or vorticity.

- We have quantified the color and the altitude-opacity indexes that serve as a diagnosis of the atmospheric level and color of the oval. It seems to be more elevated than the reference area with a value of AOI higher than 1 in all the analyzed years. From 2015, when the oval is catalogued as white, it is less red than the reference area with values of CI between $1.0\pm0.1$ and $1.06\pm0.07$, but in 2013 we obtain a $0.96\pm0.07$ value for the CI index which means it is redder than the reference but not as red as the GRS or the oval BA.

In summary, we have conducted a study of one of the longest-lived vortices in Jupiter's atmosphere over an extended period of time. Even though it would have been desirable to have a more homogeneous coverage in terms of spatial resolution and filters used, it is still possible to support the resilience of the vortex to all the environmental changes happening at the cloud level. These were external, including mergers and planetary-scale disturbances, and internal, in the form of morphology, altitude and color changes; none of them destroyed the vortex. The drift rate and vorticity of the vortex are clearly linked to its latitudinal location and they are remarkably stable in time. All these clues indicate that long-lived vortices are very likely sustained by the atmosphere at levels much deeper than the observable cloud level. However, to fully support this conclusion, more observations of this and other similar vortices are required on a long-term basis and with a temporal, spatial and spectral resolution as extended and homogeneous as possible.

## Aknowledgements

This work has been supported by the Spanish projects AYA2015-65041-P, PID2019-109467GB-100 (MINECO/FEDER, UE) and Grupos Gobierno Vasco IT1366-19. P. Iñurrigarro acknowledges a PhD scholarship from Gobierno Vasco. I. Ordonez-Etxeberria's was supported by contract from Europlanet 2024 RI. Europlanet 2024 RI has received funding from the European Union's Horizon 2020 research and innovation programme under grant agreement No 871149. This work used data acquired from the NASA/ESA HST Space Telescope, associated with OPAL program (PI: Simon, GO13937) and programs GO/DD 13067 (PI: Glenn Schneider), GO 14661 (PI: Michael Wong) and GO 14839 (PI: Imke de Pater), and archived by the Space Telescope Science Institute, which is operated by the Association of Universities for Research in Astronomy, Inc., under NASA contract NAS 5-26555. HST/OPAL maps are available at http://dx.doi.org/10.17909/T9G593. Junocam images are available at https://www.missionjuno.swri.edu/junocam/ and at the PDS Cartography and Imaging Sciences Node at: https://pds-imaging.jpl.nasa.gov/volumes/juno.html. This research has made use of the USGS Integrated Software for Imagers and Spectrometers (ISIS). PlanetCam observations were collected at the Centro Astronómico Hispánico en Andalucía (CAHA), operated jointly by the Instituto de Astrofísica de Andalucia (CSIC) and the Andalusian Universities (Junta de Andalucía) and are available on request from the instrument PI Agustín Sánchez-Lavega. Amateur images are available at the PVOL website https://pvol2.ehu.eus. LAIA and PLIA can be downloaded from: http://www.ajax.ehu.es/Software/laia.html and http://www.ajax.ehu.es/PLIA respectively. The software PICV is available at zenodo with doi: 10.5281/zenodo.4312674.

Sánchez-Lavega, A, 2011. An Introduction to Planetary Atmospheres, Taylor-Francis, CRC Press, Florida, pp. 629.

Sánchez-Lavega, A., J. Legarreta, E. García-Melendo, R. Hueso, S. Pérez-Hoyos, J. M. Gómez-Forrellad, L.N. Fletcher, G. S. Orton, A. S. Miller, N. Chanover, P. Irwin, P. Tanga, M. Cecconi, and the IOPW and ALPO Japan contributors, 2013. Colors of Jupiter's large anticyclones and the interaction of a Tropical Red Oval with the Great Red Spot in 2008. *Journal of Geophysical Research: Planets*, 118, 1-21.

Sánchez-Lavega, A., S. Pérez-Hoyos, R. Hueso, T. del Rio-Gaztelurrutia and A. Oleaga, 2014. The Aula EspaZio Gela and the Master of Space Science and Technology in the Universidad del País Vasco (University of the Basque Country). *European Journal of Engineering Education*, **39**, 516-524.

Sánchez-Lavega, A., J. H. Rogers, G. S. Orton, E. García-Melendo, J. Legarreta, F. Colas, J. L. Dauvergne, R. Hueso, J. F. Rojas, S. Pérez-Hoyos, I. Mendikoa, P. Iñurrigarro, J. M. Gomez-Forrellad, T. Momary, C. J. Hansen, G. Eichstaedt, P. Miles, and A. Wesley, 2017. A planetary-scale disturbance in the most intense Jovian atmospheric jet from JunoCam and ground-based observations. *Geophysical Research Letters*, 44, 10.

Sánchez-Lavega, A., Hueso, R., Eichstädt, G., Orton, G., Rogers, J., Hansen, C.J., Momary, T., Tabataba-Vakili, F., Bolton, S., 2018. The Rich Dynamics of Jupiter' sGreat Red Spot from JunoCam: Juno Images. *Astron. J.* 156:162 (9pp).

Sánchez-Lavega, A. et al. 2021, Jupiter's Great Red Spot: strong interactions with incoming anticyclones in 2019. *Journal of Geophysical Research,* submitted.

Simon A. A., Beebe, R. F., Gierasch, P. J., Vasavada, A. R., Belton M. J. S. and Galileo Imaging Team. 1998. Global Context of the Galileo-E6 Observations of Jupiter's White Ovals. *Icarus*, 135, 220-229.

Simon-Miller, A. A., D. Banfield, and P. J. Gierasch, 2001a. An HST study of Jovian chromophores, *Icarus*, 149, 94–106.

Simon-Miller, A. A., D. Banfield, and P. J. Gierasch, 2001b. Color and the vertical structure in Jupiter's belts, zones, and weather systems, *Icarus*, 154, 459–474.